\documentclass[11pt,twoside]{article}
\usepackage{./asp2014}

\aspSuppressVolSlug
\resetcounters

\newcommand{\beq}{\begin{equation}}
\newcommand{\eeq}{\end{equation}}
\newcommand{\dd}{{\rm{d}}}

\newcommand{\thetao}{{\theta_{\rm{o}}}}
\newcommand{\nuo}{{\nu_{\rm{o}}}}
\newcommand{\thetae}{{\theta_{\rm{e}}}}
\newcommand{\nue}{{\nu_{\rm{e}}}}
\newcommand{\re}{{r_{\rm{e}}}}
\newcommand{\rin}{{r_{\rm{i}}}}
\newcommand{\rout}{{r_{\rm{o}}}}
\newcommand{\astar}{{a_{*}}}
\newcommand{\const}{{\mbox{const}}}
\newcommand{\bmath}[1]{\mbox{\boldmath{${#1}$}}}

\begin{document}

\title{The method of transfer functions to describe GR effects in spectra and polarisation from black-hole accretion disks}
\author{Vladim\'{\i}r Karas, Michal Bursa and Michal Dov\v{c}iak
\affil{Astronomical Institute, Czech Academy of Sciences, Bo\v{c}n\'{\i} II 1401, CZ-14100~Prague, Czech Republic; \email{vladimir.karas@cuni.cz}}}
\paperauthor{Vladimir Karas}{vladimir.karas@cuni.cz}{ORCID_Or_Blank}{Astronomical Institute}{Czech Academy of Sciences}{Prague}{Praha 4, Bocni II 1401}{CZ-14100}{Czech Republic}

\begin{abstract}
We briefly review a fruitful approach to compute a variety of radiation signatures of General Relativity (GR) originating from accretion disks in strong gravity. A set of transfer functions can be pre-computed and then employed to accelerate the ray-traycing computations, generate a series of model spectra, and to fit the model predictions to actual data sets in X-rays. We have been developing this method to examine spectra and light curves and to model the expected polarimetric properties; in particular, to analyse the properties of the ``corona--disk-line'' geometry with GR effects taken into account. New impetus to this activity is emerging in the anticipation of upcoming missions equipped with the sensitivity to polarimetric properties in X-rays: IXPE (Imaging X-Ray Polarimetry Explorer) and eXTP (enhanced X-ray Timing and Polarimetry mission).
\end{abstract}

\section{Introduction} 
Over the past five decades, i.e.\ since the late 1960s, two waves of astrophysically motivated interest can be clearly identified in connection with observational effects of General Relativity in electromagnetic radiation spectra and light curves from a source embedded in strong gravitational field, in particular, an accretion disk near a black hole. First, shortly after the discovery of quasars it has been speculated that imprints of massive central black holes could be revealed in the radiation signal propagating to a distant observer. 

Idealized situations have been discussed during 1970s (Cunningham \& Bardeen 1973;  Thorne 1974), leading to the conclusion that observed radiation flux as well as the continuum (thermal) spectrum of a source must be substantially modified by the presence of supermassive black hole in the core. Characteristic features were identified in these pioneering works and the transfer function was introduced. The role of the disk self-obscuration, self-irradiation, and spectral hardening were also explored (e.g., Karas \& Bao 1992; Davis \& Laor 2011). Soon it was recognized that the origin of relativistic effects is twofold (Cunningham 1975): (i)~the structure of an accretion disk and its locally generated emission depend on the location of the innermost stable and the marginally bound orbits, which are in turn defined by the angular momentum (spin) of the black hole; and (ii)~the observable spectrum is different from its intrinsic form that is emitted in the local disk frame (LDF; local disk frame) co-moving with the gas. This is particularly true if observer's inclination is large, i.e., the disk is seen at (almost) edge-on view angle.

The accretion disk radiation (X-rays in the energy range of $0.1\lesssim E\lesssim 10^2$ keV) arises from the inner region, where the effects of GR are most profound. Complex emission features have been reported, in particular, large equivalent width of the fluorescence K$\alpha$ emission of iron; the position of its energy centroid can be understood if the iron line is produced in a medium at a state of rapid rotational motion very near a black hole (Reynolds \& Fabian 2008). Unfortunately, current models suffer mostly from an excessive number of free parameters; the observational evidence does not restrict the free parameters of accretion models in a unique manner. In the near future the observational constraints will be significantly improved by employing the additional information from X-ray polarimetry (Costa et al. 2001; De Rosa et al. 2019), but this has to proceed hand in hand with the necessary improvements in the original, ``standard'' elementary scenario of a stationary and axially symmetric flow  (cf.\ the original approaches of Cunningham 1975; Kojima 1991; Laor 1991).

\section{Overview of the adopted approach: ray tracing in Kerr geometry}
The task to determine observed line profiles originating in a source near a black hole has been solved by numerous authors using different approximations and approaches: (i)~mainly analytical scheme is made possible thanks to the remarkable property of integrability of light rays, which implies that null geodesics can be expressed via elliptic integrals in a closed form and the resulting trajectories do not exhibit chaotic behaviour; (ii)~mainly numerical ray-tracing tools are very flexible in terms of including different effects of the intervening medium that define the light transfer jointly with the geometrical properties of the curved spacetime; (iii)~a combination of the previous two ways helps to optimize the computational speed and accuracy. The latter methodology has been successfully developed in a dual-step packages, where the form geometrical effects on light rays are first pre-computed and the resulting energy shifts, light-bending amplification and rotation angles of the polarization vector are recorded in a form of a ``catalogue'' of light rays that are then used repetitively in spectral modeling by inserting different prescription for the intrinsic emissivity (Karas et al.\ 1992). 

We adopt geometrical optics approximation and employ a ray-tracing code in the Kerr black hole metric. In this context the method can be considered as an extension to Laor's (1991) study with the locally emitted intensity in the form $I_{\rm{e}}(\re,\nu)\propto\delta(\nu-\nue)J(\re)$; $\nue$ denotes emitted frequency at radial distance $\re$). Although the azimuthal velocity dominates in standard thin-disk scheme, the advection component and non-zero geometrical thickness can be taken into account, too. Effects of the disk self-gravity were also discussed (Karas et al.\ 1992).

\articlefiguretwo{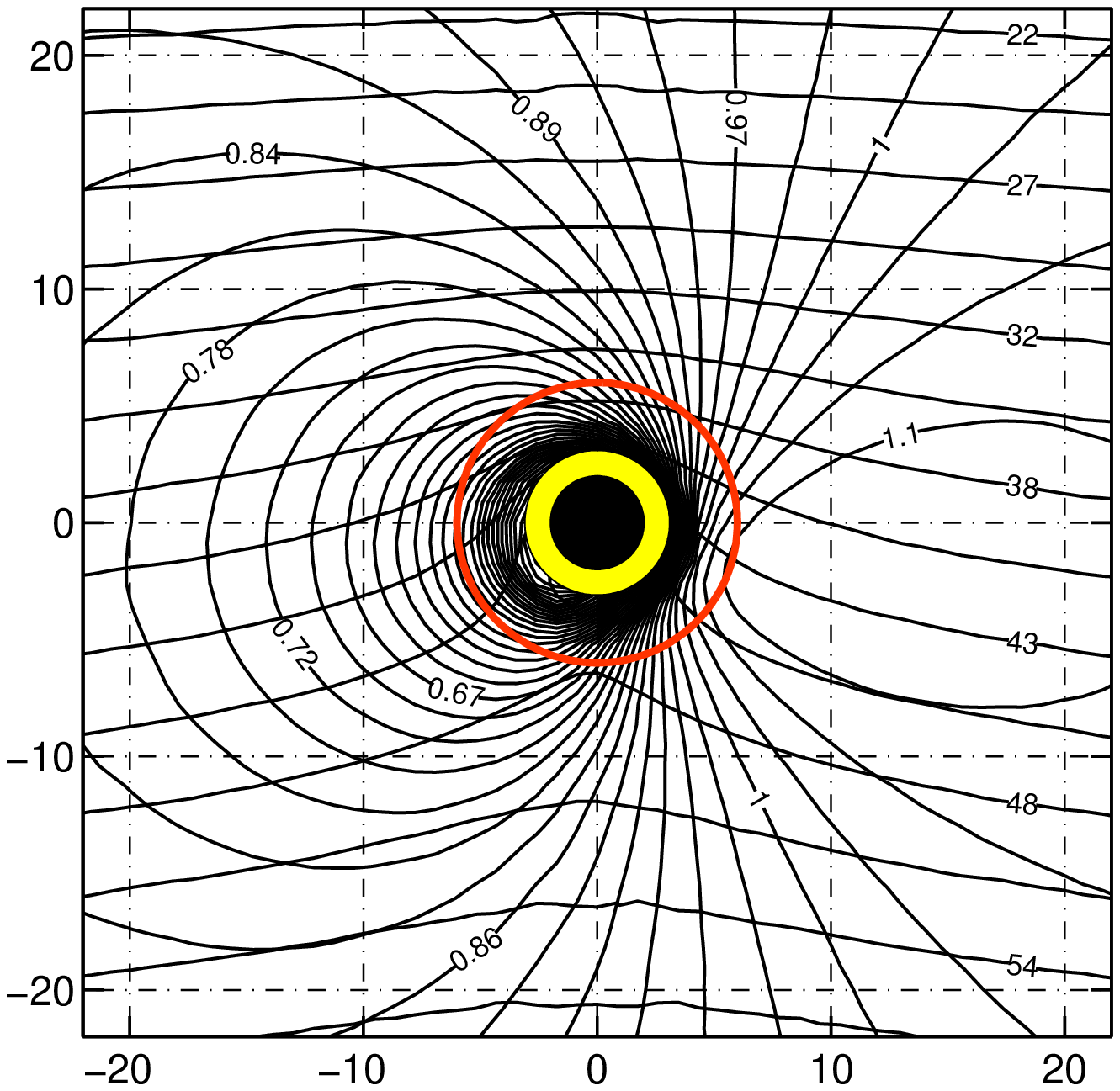}{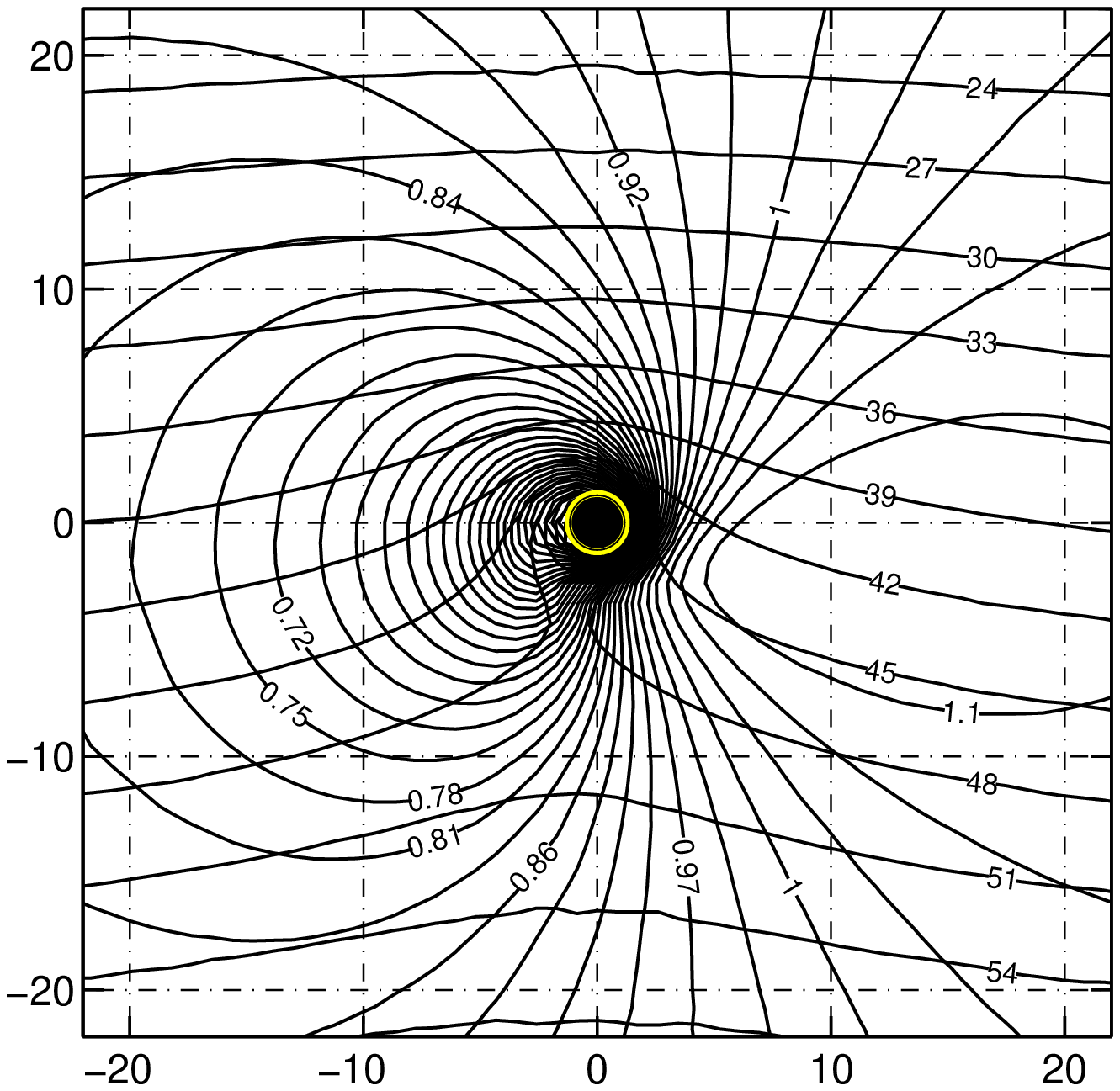}{ex_fig1}{Isocontours of constant light-travel time $\delta{t}$ and the redshift function $1+z$ within 20 gravitational radii of the equatorial plane of the Kerr black hole (black circle). These are two (of five) quantities that form the set of transfer functions to define the GR effects on light rays. \emph{Left:} the non-rotating (Schwarzschild black hole, $a=0$) case. \emph{Right:} the maximally rotating case (extreme Kerr black hole, $a=M$). This figure has been adapted from Karas (2006).}
\label{appa}
Observable (continuum subtracted) radiation flux $F_{\rm{o}}$ at frequency $\nuo$ is
\beq
 F_{\rm{o}}(\thetao,\nuo)= \int_{\cal{O}}\dd{\cal{P}}\,I_{\rm{o}}(\thetao,\nuo),
 \label{fo}
\eeq
where the integration is carried out over the observer plane, ${\cal{P}}$, which represents a detector characterized by a viewing angle $\thetao$ and located in the asymptotically flat radial infinity, $r \gg GM/c^2$. Integration in eq.~(\ref{fo}) can be rewritten in terms of quantities on the disk surface, ${\cal{S}}$,
\beq
 F_{\rm{o}}(\thetao,\nuo)=
 \int_{\cal{S}}\dd{\cal{S}}\int_{g_{\rm{min}}}^{g_{\rm{max}}}\dd{g}\,
 I_{\rm{e}}(\re,\thetae,\nue)T(\re,\thetae,\nue;\thetao,\nuo).
 \label{fo1}
\eeq
Here, the transfer function $T$ determines what fraction of locally emitted energy reaches the observer when only geomtrical factors are taken into account, while the locally
emitted intensity is related to observed intensity by $I_{\rm{e}}(\re,\nue)/\nue^3=I_{\rm{o}}(\nuo)/\nuo^3$ ($g=\nuo/\nue$ denotes the overall, Doppler plus gravitational redshift factor). Instead of explicit calculation of the transfer function according to eq.~(\ref{fo1}) one can equivalently divide the disk image in individual pixels and evaluate contributions to the total radiation flux directly at infinity, as in eq.~(\ref{fo}).

Light rays connect points of emission on the disk surface with the corresponding pixels in the detector plane. As mentioned above, the light rays are represented by null geodesics in the Kerr spacetime. Again, a variety of approaches have been developed (e.g. Bromley et al.\ 1997) and they are equivalent in principle but very much different in practical implementations. Karas et al.\ (1992) found it useful to approximate an extensive set of integrated rays by Chebyshev polynomials. Each photon ray in Kerr metric is defined by its specific angular momentum $\xi$ and Carter's constant $\eta$. 

\articlefiguretwo{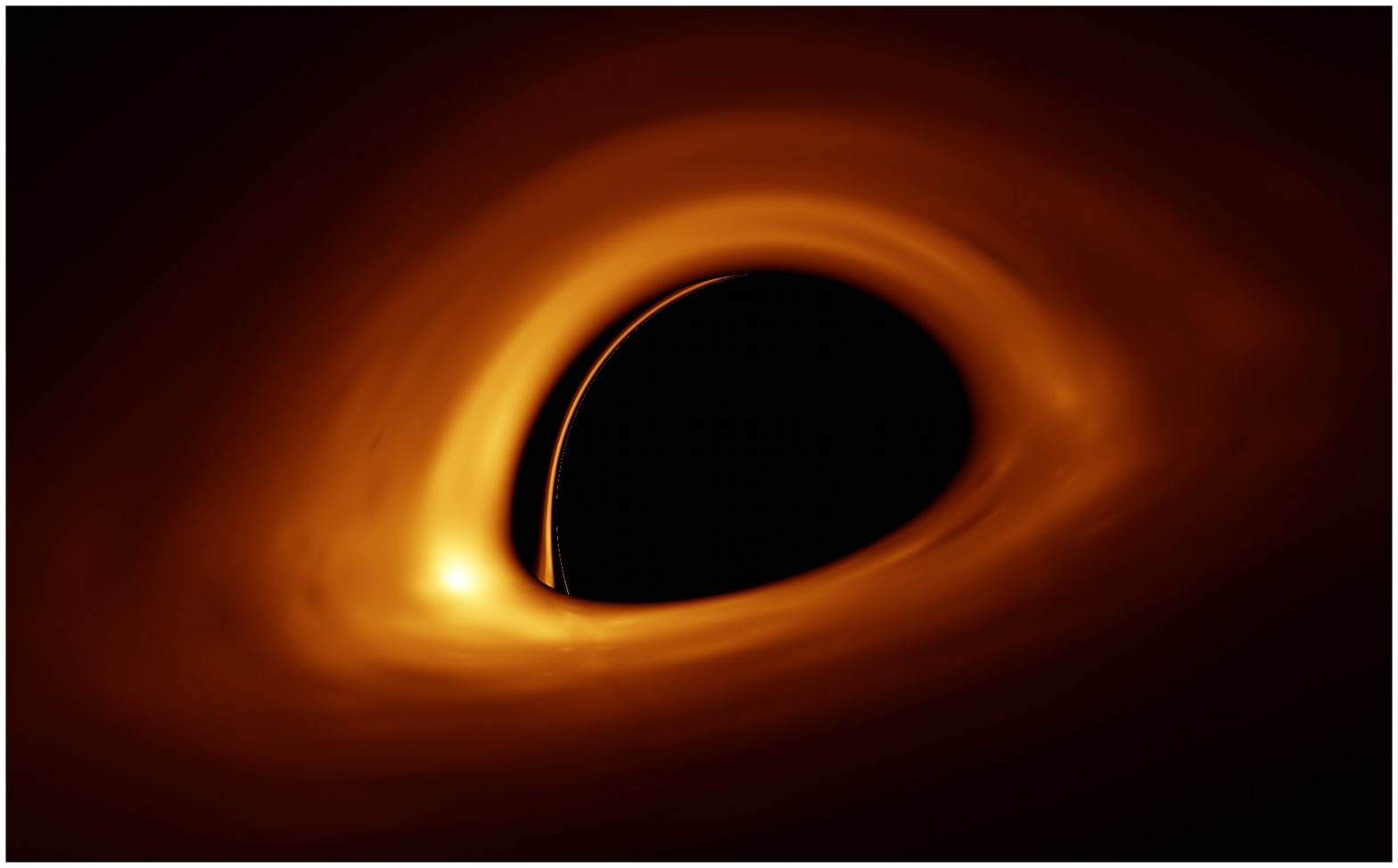}{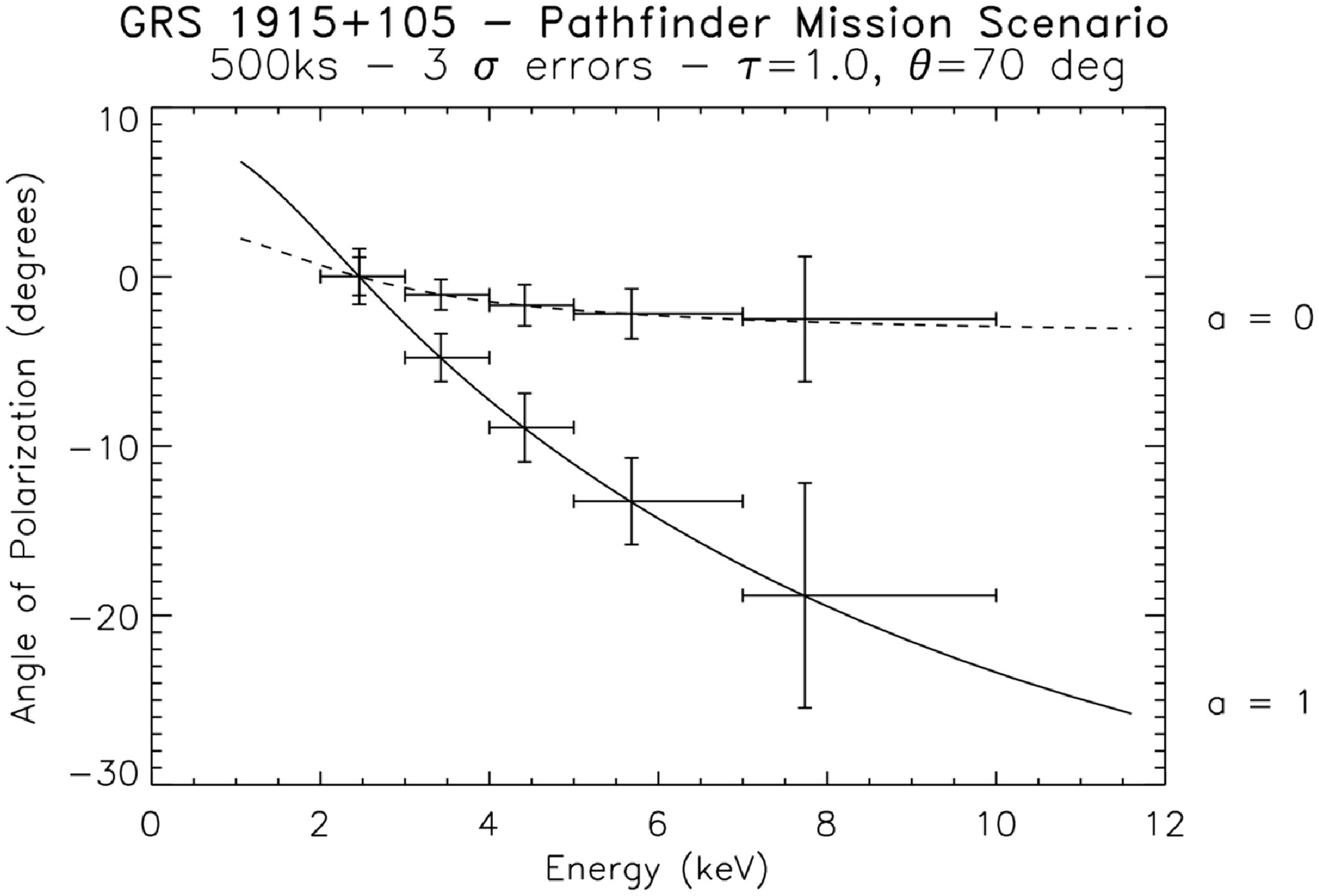}{ex_fig2}{The inner region of a standard thin accretion disk orbiting around a black hole. \emph{Left:} A distorted image near the innermost stable circular orbit and the light circle are shown (Bursa et al.\ 2007). \emph{Right:} The model energy dependence of the polarization angle from the thermal disk emission, as predicted for 500 ks pathfinder mission of the bright microquasar (Dov\v{c}iak et al.\ 2008).}

Wildly bent rays connect the points of emission from the black hole disk plane to the observer's plane, ${\cal{P}}$; in polar coordinates ${\rho},$ ${\varphi}$: $\xi={\rho}\sin\theta_0\sin{\varphi}$, $\eta={\rho}^2\cos^2{\varphi}-{\astar}^2\cos^2\theta_0+\xi^2\cot^2\theta_0$. The geodesic equation relates $\re$ and $\thetae$ to the final value 
$\thetao$ of the outgoing photons at infinity (standard notation for the Kerr geometry in Boyer-Lindquist coordinates and dimensionless geometric units),
\beq
 \int_\infty^{\re} {\dd{r}\over{}R(r;\xi,\eta,{\astar})^{1/2}} =
 \int_{\cos\thetao}^{\cos\thetae}
  {\dd\cos\theta\over\Theta(\cos\theta;\xi,\eta,{\astar})^{1/2}},
 \label{carter}
\eeq
where the integrands are given by polynomials in separated variables $r$ and $\theta$,
\begin{eqnarray*}
 R(r;\xi,\eta,{\astar}) & = & r^4 + \left({\astar}^2 - \xi^2 -
 \eta\right)r^2 + 2\left[\eta
 + \left(\xi-{\astar}\right)^2\right]r - {\astar}^2\eta,
 \label{rfun} \\
 \Theta(\theta;\xi,\eta,{\astar}) & = &\eta-\left(\xi^2+\eta-
 {\astar}^2\right)\cos^2\theta-{\astar}^2\cos^2\theta.
 \label{thfun}
\end{eqnarray*}
The relation $\re\equiv\re(\thetae)$ gives the disk photosphere, which stands as an input of the computation. The point of emission can be expressed from eq.~(\ref{carter}) in terms of Jacobian elliptic functions of parameters $\thetao$, $\xi$, $\eta$ (Byrd \& Friedman 1971). We then solve eq.~(\ref{carter}) for $\re(\thetae;\thetao)$ to find the form of integrals (\ref{carter}) for different roots in eq.\ (\ref{rfun}). 

Starting from above-given definition of the redshift function $g$, one can derive its explicit form in equatorial plane of the Kerr metric $g_{\alpha\beta}$. Taking into account both the azimuthal and radial motion we find
\begin{equation}
g  =  \frac{p_\alpha\eta^\alpha}{\tilde{p}_\alpha\tilde{\eta}^\alpha}
    =  \left[g^{tt}\tilde{\eta}_t
   +g^{\phi{t}}\left(\tilde{\eta}_\phi-\xi\tilde{\eta}_t\right)
   -g^{\phi\phi}\xi\tilde{\eta}_\phi+
   +g^{rr}\tilde{\eta}_r\tilde{p}_r/p_t\right]^{-1},
  \label{gfun}
\end{equation}
where $p^\alpha$ and $\eta^\alpha$ denote four-momentum of the detected photon and the observer at $r\rightarrow\infty$, respectively, and analogously $\tilde{p}^\alpha$ and $\tilde{\eta}^\alpha$ for photon and the emitting material locally in the disk. 

In terms of Boyer-Lindquist coordinates, $\Delta=r^2-2r+{\astar}^2$, $\Sigma=r^2+{\astar}^2\cos^2\theta$, $A=(r^2+{\astar}^2)^2-\Delta{{\astar}^2}\sin^2\theta$. For the five terms in brackets on r.h.s.\ of eq.~(\ref{gfun}),
\begin{eqnarray}
 g^{tt}\tilde{\eta}_t & = &
  \left(\sqrt{\frac{A}{\Sigma\Delta}}
  +\frac{2{\astar}r\sqrt{A}}{\Sigma^{3/2}\Delta}\,v^{(\phi)}\right)\gamma,
  \quad
 g^{t\phi}\tilde{\eta}_\phi =
  -\frac{2{\astar}r\sqrt{A}}{\Sigma^{3/2}\Delta}\,\gamma{v}^{(\phi)},
  \\
 g^{\phi{t}}\xi\tilde{\eta}_t & = &
  -2{\astar}r\left[\frac{1}{\sqrt{\Sigma\Delta{A}}}
  +\frac{2{\astar}r}{\Sigma^{3/2}\sqrt{A}\Delta}\,v^{(\phi)}
  \right]\gamma\xi,
  \\
 g^{\phi\phi}\xi\tilde{\eta}_\phi & = &
  \frac{\left(r^2-2r\right)\sqrt{A}}{\Sigma^{3/2}\Delta}\,\gamma
  {v}^{(\phi)}\xi,
  \quad
 g^{rr}\tilde{\eta}_r\frac{\tilde{p}_r}{p_t}  = 
  \frac{\sqrt{\Delta{R}}}{\Sigma^{3/2}}\,\gamma{v}^{(r)}\sigma_r.
\end{eqnarray}
Here, $v^{(\phi)}$, $v^{(r)}$ are azimuthal and radial components of material orbital velocity $\bmath{v}$, $\sigma_r$ is signature of $p^r\propto\sigma_r\sqrt{R}$, and $\gamma=1/\sqrt{1-v^2}$. Finally, the intrinsic angle of emission $\vartheta$ between photon and normal to the disk surface is
$\cos\vartheta=-(\bmath{p\cdot{n}})/(\bmath{p\cdot\eta})$.

\articlefiguretwo{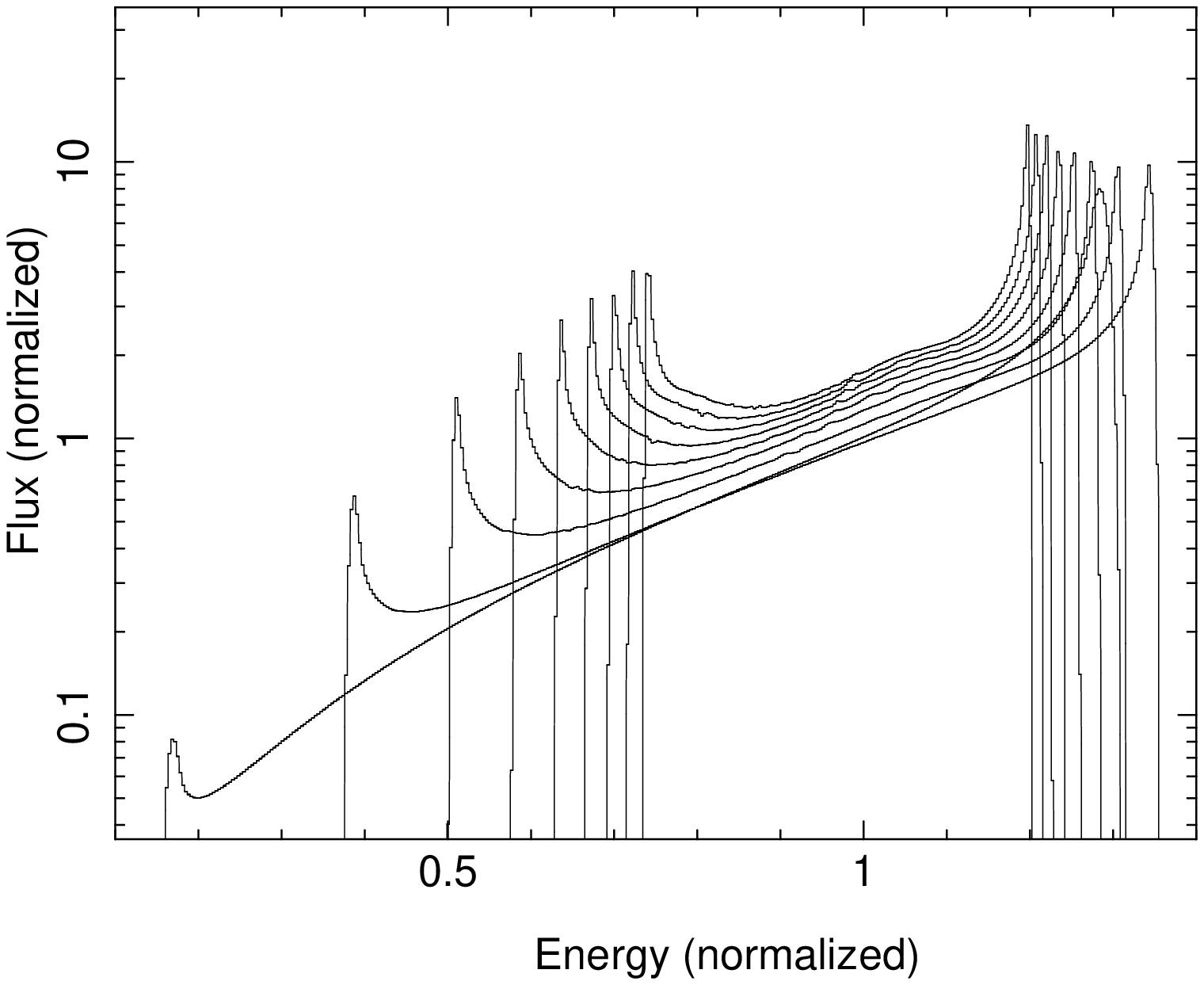}{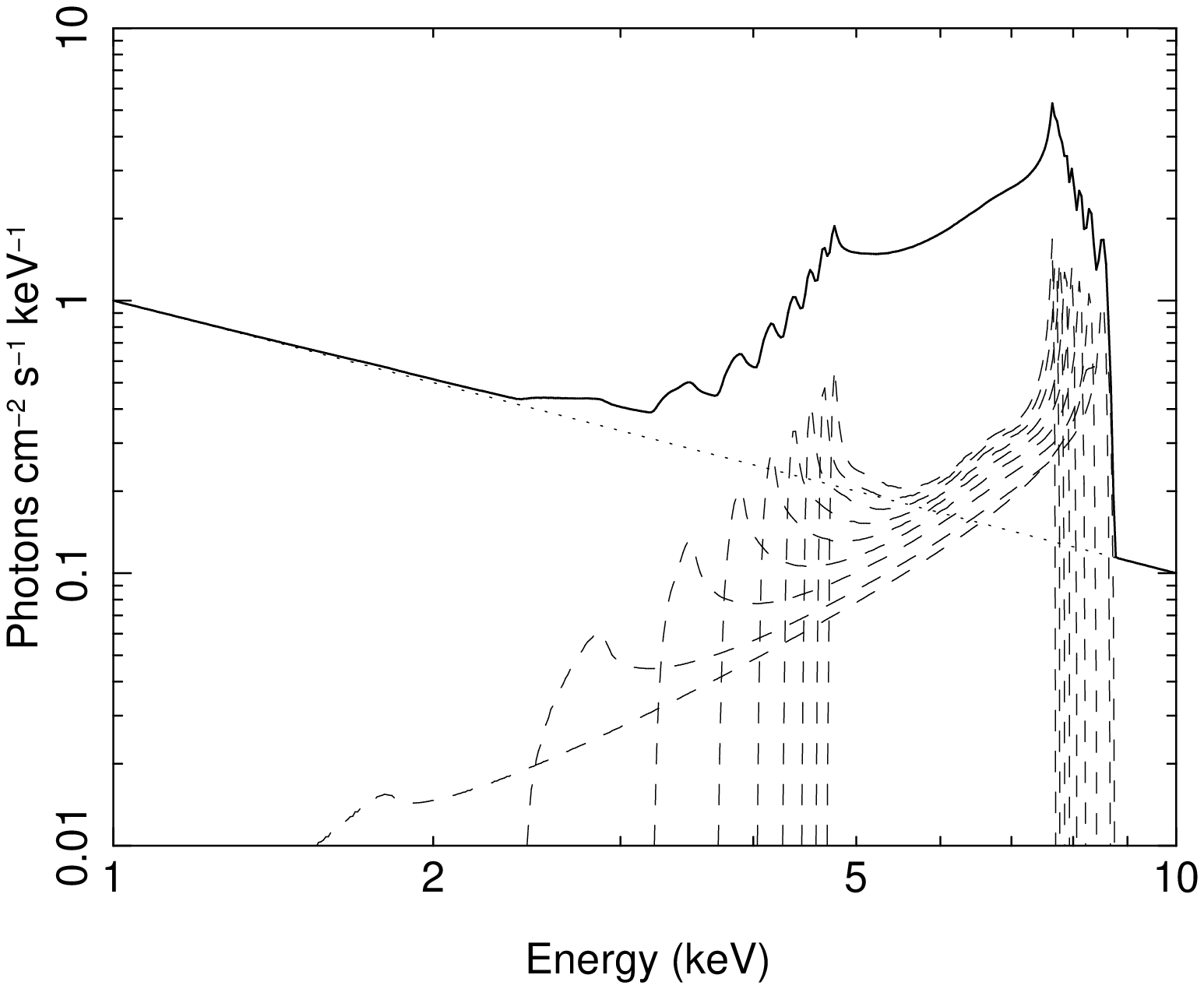}{ex_fig3}{A relativistically broadened and skewed profile of a spectral line has been foreseen to emerge from a black-hole accretion disk. The position, relative intensity and the equivalent width offer a way to measure the parameters. Left: the theoretical profiles (arbitrary units, continuum subtracted) from a set of nine infinitesimally narrow rings in the equatorial plane of a Kerr black hole. Right: the rest energy of the line has been set to $6.4$~keV and a power-law continuum added. Dashed lines denote the individual components. The signature of the individual rings is visible in the wings of the final profile (Karas \& Sochora 2010).}

As a particular example (often used in previous works), the assumption of purely azimuthal Keplerian motion leads to
\begin{equation}
 v^{(\phi)}  =  \frac{r^2-2a\sqrt{r}+{\astar}^2}{\sqrt{\Delta}\left(r^{3/2}
 +{\astar}\right)},\quad
 \gamma  =  \frac{\left(r^{3/2}+{\astar}\right)\sqrt{\Delta}}{r^{1/4}
  \sqrt{r^{3/2}-3r^{1/2}+2{\astar}}\sqrt{r^3+{\astar}^2r+2{\astar}^2}}.
\end{equation}
The corresponding angular velocity of orbital motion is the $\Omega\equiv\eta^\phi/\eta^t=1/(r^{3/2}+{\astar})$, and the redshift function and the emission angle are 
\begin{equation}
 g = \frac{{\cal{C}}}{{\cal{B}}-r^{-3/2}\xi}, \quad \vartheta  =  \arccos\frac{g\sqrt{\eta}}{r}
\end{equation}
with ${\cal{B}}=1+{\astar}r^{-3/2}$, ${\cal{C}}=1-3r^{-1}+2{\astar}r^{-3/2}$. In the case of the standard Keplerian disk, the local energy dissipation due to viscous effects produces the radiation flux
\beq
 F(r)\propto
 \frac{\partial\Omega}{\partial{r}}
 \left(\eta_t+\Omega\eta_\phi\right)^{-2}
 \int_{\rin}^{\re}\left(\eta_t+\Omega\eta_\phi\right)
 \frac{\partial\eta_\phi}{\partial{r}}{\dd}r
 \label{ntemis}
\eeq
 (Page \& Thorne 1974). Fig.\ \ref{ex_fig1} shows an exemplary graphical representation of the transfer function structure along the the equatorial plane. See Karas (2006) and Ranea-Sandoval \& Garc\'{\i}a (2015) for further details and references.

\section{Results}
\label{results}
The disk is characterized by radiation intensity $I_{\rm{em}}$ emitted locally from its surface, $z\equiv{z(\re)}$ at frequency $\nue$, $I_{\rm{em}}(\nue;\re,\thetae)=F(\re)\varphi_1(\nue)\varphi_2(\vartheta)$, where $F(\re)$ is the total radiation flux at the disk surface, $\varphi_1(\nue)$ is the emissivity profile in frequency,
$\varphi_2(\vartheta)$ is the limb-darkening law; $\vartheta$ the angle between the ray and direction normal to the disk. 

We can classify different models by the following parametrization of the profiles of intrinsic emissivity (Dov\v{c}iak et al. 2004; Caballero-Garc\'{\i}a et al. 2018):
(i)~Geometrical shape: height of the disk (function $z(\re)$; $z=0$ for an equatorial disk) and its size (inner edge $\rin$, outer edge $\rout$);
(ii)~Local radiation flux, $F(\re)$. To this end we examined, e.g., the power-law dependence $F(\re)\propto{\re}^{-\beta}$ ($\beta=\const$), and the standard  Novikov--Thorne--Page (Dewit \& Dewitt 1973) scheme for the geometrically thin accretion regime, as described by eq.\ (\ref{ntemis});
(iii)~Emissivity profile in frequency, $\varphi_1(\nue)$. We examined Gaussian profiles $\varphi_1(\nue)\propto\exp[-\varepsilon(\nue-1)]^2$ ($\varepsilon=\const$), and asymmetric profiles that account for effects of self-irradiation;
(iv)~Angular dependence of the local emissivity (limb-darkening law), $\varphi_2(\vartheta)$. We examined, for example, exemplary cases of  $\varphi_2(\vartheta)=1+\epsilon\mu$ ($\mu=\cos\theta$, $\epsilon=\const$); and $\varphi_2(\vartheta)=\mu\log(1+1/\mu)$.

See Fig.\ \ref{ex_fig2} for a synthetic image of a black hole accretion disk. The relativistic effects deform its apparent shape and they even produce indirect images that are formed by photons circling multiple times. The predicted polarization gives prospects for future ways of constraining the black hole spin. Figure \ref{ex_fig3} shows a typical double-horn spectral line that arises by superposing profiles from several narrow accretion rings, eventually forming a radially extended Keplerian disk (radius of the rings increases gradually from $r=2$ to $r=18$ gravitational radii). The parameters are: observer's view-angle inclination $75^{\circ}$ (close to edge-on line of sight) and the black hole spin $a=0.6$ (moderately fast prograde rotation along the black hole angular momentum). The latter parameters of the system can be then measured by fitting the observed spectrum to the model. Finally, let us note that more complicated scenarios can be also adopted: for example, the effects of accretion disk self-gravity (Karas et al. 1995) or non-negligible geometrical thickness of the accretion flow (Taylor \& Reynolds 2018).

\acknowledgements The authors acknowledge the initial support from the Czech Ministry of Education, Youth and Sports program Kontakt (LTAUSA 17095), and the Czech Science Foundation -- DFG collaboration project (ref.\ 19-01137J).


\begin{thebibliography}{}
\bibitem[Bromley et al.(1996)]{BC97} Bromley B.~C., Chen~K., \& Miller W.~A. 1997, ApJ 475, 57
\bibitem[Bursa et al.(2007)]{B07} Bursa M., Abramowicz M. A., Karas V., et al.\ 2007, in {\it RAGtime 8/9: Workshops on Black Holes and Neutron Stars}, eds.\ S.\ Hled\'{\i}k \& Z.\ Stuchl\'{\i}k (Opava: Silesian University),  pp. 21-25
\bibitem[Byrd \& Friedman(1971)]{BF71} Byrd P.~F., \& Friedman M.~D. 1971, {\it Handbook of Elliptic Integrals for Engineers and Scientists} (Berlin: Springer-Verlag)
\bibitem[Caballero-Garcia et al.(2018)]{CG18} Caballero-Garc\'{\i}a M. D., Papadakis I. E., Dov\v{c}iak M., Bursa M., et al. 2018, MNRAS 480, 2650
\bibitem[Costa et al.(2001)]{C01} Costa E., Soffitta P., Bellazzini R. et al. 2001, Nature 411, 662
\bibitem[Cunningham(1975)]{C75} Cunningham C.~T. 1975, ApJ 202, 788
\bibitem[Davis \& Laor(2011)]{DL11} Davis S. W., \& Laor A. 2011, ApJ 728, id. 98
\bibitem[Dewitt \& Dewitt(1973)]{DD73} Dewitt C., \& Dewitt B. S. (eds.) 1973, {\it Black Holes} (New York: Gordon and Breach)
\bibitem[De Rosa et a.(2019)]{DR19} De Rosa A., Uttley P., Gou Lijun, Liu Yuan, et al.\ 2019, Sci. China -- Phys. Mech. Astron. 62(2), 029504 (arXiv:1812.04022)
\bibitem[Dov\v{c}iak et a.(2004)]{D04} Dov\v{c}iak M., Karas V., \& Yaqoob T. 2004, ApJSS 153, 205
\bibitem[Dov\v{c}iak et a.(2008)]{D08} Dov\v{c}iak M., Muleri F., Goosmann R. W., Karas V., \& Matt G. 2008, MNRAS 391, 32
\bibitem[Karas(2006)]{K06}Karas V. 2006, AN 327, 961
\bibitem[Karas \& Bao(1992)]{KB92} Karas V., \& Bao G. 1992, A\&A 257, 531
\bibitem[Karas et al.(1995)]{KLV95} Karas V., Lanza A., \& Vokrouhlick\'y 1995, ApJ 440, 108
\bibitem[Karas \& Sochora(2010)]{KS10}Karas V., \& Sochora V. 2010, ApJ 725, 1507
\bibitem[Karas et al.(1992)]{K92}Karas V., Vokrouhlick\'y D., \& Polnarev A. 1992, MNRAS 259, 569
\bibitem[Kojima(1991)]{K91}Kojima Y. 1991, MNRAS 250, 629
\bibitem[Laor(1991)]{L91} Laor A. 1991, ApJ 376, 90
\bibitem[Page \& Thorne(1974)]{PT74}Page D. N., \& Thorne K. S. 1974, ApJ 499, 191
\bibitem[Ranea-Sandoval \& Garc\'{\i}a(2015)]{RG15}Ranea-Sandoval I. F., \& Garc\'{\i}a F. 2015, A\&A 574, id. A40
\bibitem[Reynolds \& Fabian(2008)]{RF08} Reynolds C. S., \& Fabian A. C. 2008, ApJ 675, 1048
\bibitem[Taylor \& Reynolds(2018)]{TR18} Taylor C., \& Reynolds C. S. 2018, ApJ 868, id. 109
\end{thebibliography}
\end{document}